\documentclass[conference, a4paper]{IEEEtran}
\usepackage[noadjust]{cite}
\usepackage[utf8]{inputenc}
\usepackage{amsmath}
\usepackage{amsfonts,amssymb}
\usepackage{array}
\usepackage{graphicx}
\usepackage{booktabs}
\usepackage{afterpage}
\usepackage{float}
\usepackage{mathtools}
\usepackage{color}
\usepackage[thmmarks,amsmath]{ntheorem}
\usepackage[linesnumbered,ruled]{algorithm2e}
\SetKwComment{Comment}{$\triangleright$ }{}
\usepackage[numbers,sort&compress]{natbib}
\usepackage{balance}

\def\C{{\mathbb{C}}}

\def\argmax{\mathop{\mathrm{argmax}}}

\begin{document}
\setlength{\abovedisplayskip}{3pt}
\setlength{\belowdisplayskip}{3pt}
\title{
Low-Complexity Near-Field Channel Estimation for Hybrid RIS Assisted Systems\\
\author{\IEEEauthorblockN{Rafaela~Schroeder\IEEEauthorrefmark{1},
Jiguang~He\IEEEauthorrefmark{2}\IEEEauthorrefmark{1}, Hamza~Djelouat\IEEEauthorrefmark{1} and Markku~Juntti\IEEEauthorrefmark{1}}
\IEEEauthorblockA{
    \IEEEauthorrefmark{1}Centre for Wireless Communications, FI-90014, University of Oulu, Finland \\
    \IEEEauthorrefmark{2}Technology Innovation Institute, 9639 Masdar City, Abu Dhabi, United Arab Emirates}
\IEEEauthorblockA{E-mail: \{rafaela.schroeder, hamza.djelouat, markku.juntti\}@oulu.fi, jiguang.he@tii.ae}} 
}
\maketitle
\begin{abstract} 
We investigate the channel estimation (CE) problem for hybrid RIS assisted systems and focus on the near-field (NF) regime. Different from their far-field counterparts, NF channels possess a \textit{block-sparsity} property, which is leveraged in the two developed CE algorithms: (i) boundary estimation and sub-vector recovery (BESVR) and (ii) linear total variation regularization (TVR). In addition, we adopt the alternating direction method of multipliers to reduce their computational complexity. Numerical results show that the linear TVR algorithm outperforms the chosen baseline schemes in terms of normalized mean square error in the high signal-to-noise ratio regime while the BESVR algorithm achieves comparable  performance to the baseline schemes but with the added advantage of minimal CPU time.
\end{abstract}
\section{Introduction}
Reconfigurable intelligent surfaces (RISs) have emerged as a promising technology in wireless communications systems~\cite{huang2019reconfigurable}. RISs are usually constructed as planar arrays comprising numerous cost-effective radiating elements with the aid of controllable phase shifters. By adjusting these phase shifters, the RISs can modify their response to the incident signal to achieve a certain objective, such as focusing the signal towards the receiver. To reap the benefits of the RIS, high accuracy channel state information (CSI) is essential, the acquisition of which is a challenging task due to the large number of passive RIS elements.

To ease CSI acquisition, a \textit{hybrid} RIS architecture with a mixture of both passive and active receive elements can be introduced to the communication system~\cite{ActElements}. With the introduction of active RIS elements, the RIS is able to demodulate the received pilot signals during the training procedure. Several studies have been conducted to investigate the channel estimation (CE) for hybrid RIS-assisted systems~\cite{yang2022separate,h_ris_ce_bjornson,schroeder2021two}, but the majority have focused on the far-field (FF) regime, assuming a planar wavefront. In practical scenarios, the RIS may consist of a considerable number of reflective elements, extending the \textit{Rayleigh distance}, known as the boundary distinguishing the FF and near-field (NF) regimes, further from the RIS. Consequently, it is likely that the mobile station (MS) is situated in the NF of the RIS~\cite{gan2022near}. In such circumstances, the channel model follows the spherical-wavefront assumption, incorporating angles and distances. Only a limited number of recent works have studied the NF CE problem~\cite{cui2022near,wei2021channel,zhang2023near2}. For instance, Cui and Dai in~\cite{cui2022near} proposed a sparse polar-domain representation of the NF channel and estimated it via compressive sensing (CS) techniques. A significant challenge in the NF CE is the sparsity pattern in the angular domain. Unlike the FF channel, the sparsity of the NF channel in the angular domain is not concentrated in a single spike but multiple consecutive spikes. Hence, the angular-domain amplitudes of NF channels, characterized by a block of non-zero elements, can be leveraged in CE algorithm design. 

This paper introduces a novel \textit{block-sparsity} formulation for NF CE in hybrid RIS-assisted multiple-input single-output (MISO) systems. 
To tackle this problem, we propose two algorithms:~(i) boundary estimation and sub-vector recovery (BESVR) and~(ii) linear total variation regularization (TVR). To ease the complexity of the high-dimensional NF CE algorithms, we resort to the alternating direction method of multipliers (ADMM). The performance of the proposed schemes is evaluated in terms of normalized mean square error (NMSE) and average CPU time. The simulation results indicate that the TVR algorithm is able to outperform the baseline schemes in high signal-to-noise ratio (SNR). In contrast, the BESVR algorithm achieves similar NMSE performance compared to the baseline schemes with minimal CPU time. 
\begin{figure}
    \centering
    \includegraphics[width=0.6\columnwidth]{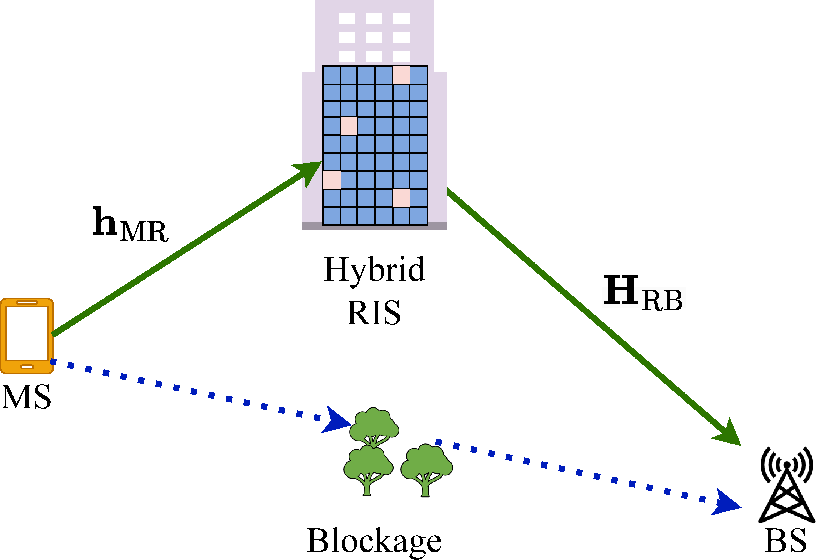}
    \caption{The considered hybrid RIS-assisted MISO systems.}
    \label{fig:hris_nf}
    \vspace{-0.5cm}
\end{figure}
\section{System Model}
The system model incorporates a hybrid RIS-aided MISO systems, comprising a multi-antenna base station (BS), a multi-element RIS with a uniform linear array (ULA), and a single-antenna MS, as depicted in Fig.~\ref{fig:hris_nf}\footnote{We offer an example of two-dimensional RIS-assisted networks in Fig.~\ref{fig:hris_nf} for better illustration propose. However, our work is focused on the ULA antenna. The extension to uniform planar array (UPA) is feasible.}. Both the BS and MS are assumed to be in the NF of the RIS. We assume perfect knowledge of the channel between the RIS and the BS, i.e, $\mathbf{H}_\text{RB}\in\mathbb{C}^{N_{\text{B}}\times N_{\text{R}}}$, where $N_{\text{B}}$ denotes the number of antennas at the BS and $N_{\text{R}}$ denotes the number of RIS elements. Thus, our focus is on estimating the NF channel vector $\mathbf{h}_{\text{MR}}\in\mathbb{C}^{N_{\text{R}}\times 1}$ between the MS and the RIS.
\subsection{Near-Field Channel Model}
Under the spherical wavefront assumption, the multi-path NF channel is formulated as~\cite{cui2022near}
\begin{small}
    \begin{equation}
    \label{eq:near_field_channel}
   \mathbf{h}_\text{MR} = \sum\limits_{l = 1}^{L_\text{MR}}\boldsymbol{\rho}_{\text{MR},l}\boldsymbol{\alpha}\big(\boldsymbol{\phi}_{\text{MR},l},r_{\text{MR},l}\big),
    \end{equation}
\end{small}
where $L_\text{MR}$ is the number of paths. The terms $\boldsymbol{\rho}_{\text{MR},l}$, $\boldsymbol{\phi}_{\text{MR},l}$, and $r_{\text{MR},l}$ are the complex path gain, the angle, and the distance associated with the $l$th path, respectively. The steering vector $\boldsymbol{\alpha}\big(\boldsymbol{\phi}_{\text{MR},l},r_{\text{MR},l}\big)$ is given as
\begin{small}
    \begin{equation}
      \label{eq:nf_svector}       \boldsymbol{\alpha}\big(\boldsymbol{\phi}_{\text{MR},l},r_{\text{MR},l}\big)\hspace{-0.1cm} =\hspace{-0.1cm}  \big[1,e^{-jk(r_{l}^{(0)}-r_{\text{MR},l})},\hspace{-0.1cm} \cdots\hspace{-0.1cm}, e^{-jk(r_{l}^{(N_\text{R}-1)}-r_{\text{MR},l})}\big]^\mathsf{T}, 
    \end{equation}
\end{small}
where $\kappa=\frac{2\pi f_c}{c}$ is the wavenumber with $f_c$ denoting the frequency and $c$ being the speed of the light. We assume that the NF channel in~\eqref{eq:near_field_channel} consists of one line-of-sight (LOS) path and one non-line-of-sight (NLOS) path. For simplicity, we assign $l=1$ for the LOS path and $l=2$ for the NLOS path. Therefore, $r^{(n)}_{1}$ represents the distances between the $n$th RIS element and the MS. Similarly, $r^{(n)}_{2}$ is the distance between the $n$th RIS element and the scatter.  
Assuming that the coordinate of the $n$th element is~$(0,\delta_{n}d)$, where $\delta_n = \frac{2n-N_\text{R}+1}{2}$, with $n=0,1,\cdots,N_\text{R}-1$, $r^{(n)}_{l}$ can be expressed as $r^{(n)}_{l} = \sqrt{r_{\text{MR},l}+\delta_{n}^{2}d^{2}-2r_{\text{MR},l}\sin(\boldsymbol{\phi}_{\text{MR},l})\delta_{n}d}$, where $d$ is the antenna inter-element spacing.
\subsection{Uplink Training Procedure}
We assume $M$ out of $N_\text{R}$ RIS elements are active and $N_\text{RF,R} < M$ RF chains are deployed at RIS. Considering that the MS sends the pilot signal $s_t$ with power $\mathbb{E}(|s_t|^2) = P$ during the $t$th symbol duration, the received signal at the hybrid RIS is expressed as
\begin{small}
    \begin{equation}
    \label{eq:received_signal_hris}
    \mathbf{y}_{\text{H},t} = \mathbf{W}_{t}\mathbf{M}_{t}\mathbf{h}_\text{MR}s_t+\mathbf{W}_{t}\mathbf{M}_{t}{\mathbf{n}_{t}},
    \end{equation}
\end{small}
where $\mathbf{M}_t$ is a row-selection matrix containing $M$ rows of a identity matrix ${N_\text{R}\times N_\text{R}}$ and $\mathbf{W}_t\in \mathbb{C}^{N_\text{RF,R}\times M}$ is the analog combining matrix\footnote{The analog combining matrix is introduced based on the assumption that the number of RF chains is smaller than the number of active elements, i.e, $N_\text{RF,R} < M$. However, for $N_\text{RF,R} = M$, the term $\mathbf{W}_t$ can be omitted in~\eqref{eq:received_signal_hris}.}. The additive white Gaussian noise (AWGN) is denoted by $\mathbf{n}_{t}\in \mathbb{C}^{N_{\text{R}}\times 1}$ with each entry distributed as $\mathcal{CN} (0,\sigma^2)$, where $\sigma^2$ denotes its variance. By collecting the received signal across $T$ time slots and setting $s_T$ to $\sqrt{P}$, the following expression is obtained 
\begin{small}
    \begin{equation}
    \label{eq:c_rsignal}
    \mathbf{y}_{\text{H}} = \sqrt{P}\mathbf{W}_\text{M}\mathbf{h}_\text{MR} + \Tilde{\mathbf{n}},
    \end{equation}
\end{small}
where $\mathbf{W}_\text{M} = [(\mathbf{W}_{1}\mathbf{M}_{1})^\mathsf{T}, \cdots,(\mathbf{W}_{T}\mathbf{M}_{T})^\mathsf{T}]^\mathsf{T} \in \mathbb{C}^{T N_\text{RF,R} \times N_\text{R}}$ and $\Tilde{\mathbf{n}} = [(\mathbf{W}_{1}\mathbf{M}_{1}\mathbf{n}_{1})^\mathsf{T}, \cdots,(\mathbf{W}_{T}\mathbf{M}_{T}\mathbf{n}_{T})^\mathsf{T}]^\mathsf{T}\in \mathbb{C}^{ T N_\text{RF,R} \times 1}$. By introducing an angular-domain transformation matrix $\mathbf{F}\in\mathbb{C}^{N_{\text{R}}\times N_{\text{R}}}$, we can rewrite the received signal in~\eqref{eq:c_rsignal} as 
\begin{small}
\begin{equation}
        \label{eq:r_final}
        \mathbf{y}_{\text{H}} = \sqrt{P}\boldsymbol{\Psi}\mathbf{z} + \Tilde{\mathbf{n}}, 
    \end{equation}    
\end{small}
where $\boldsymbol{\Psi}=\mathbf{W}_\text{M}\mathbf{F}\in\mathbb{C}^{T N_\text{RF,R} \times N_{\text{R}}}$ and $\mathbf{z}\in\mathbb{C}^{N_{\text{R}}\times1}$ is the transformation coefficient vector of the NF channel in the angular domain. The transformation matrix $\mathbf{F}$ is modeled as a discrete Fourier transform (DFT) matrix.

\begin{figure}
    \vspace{-0.3cm}
    \centering
    \includegraphics[width=0.6\columnwidth]{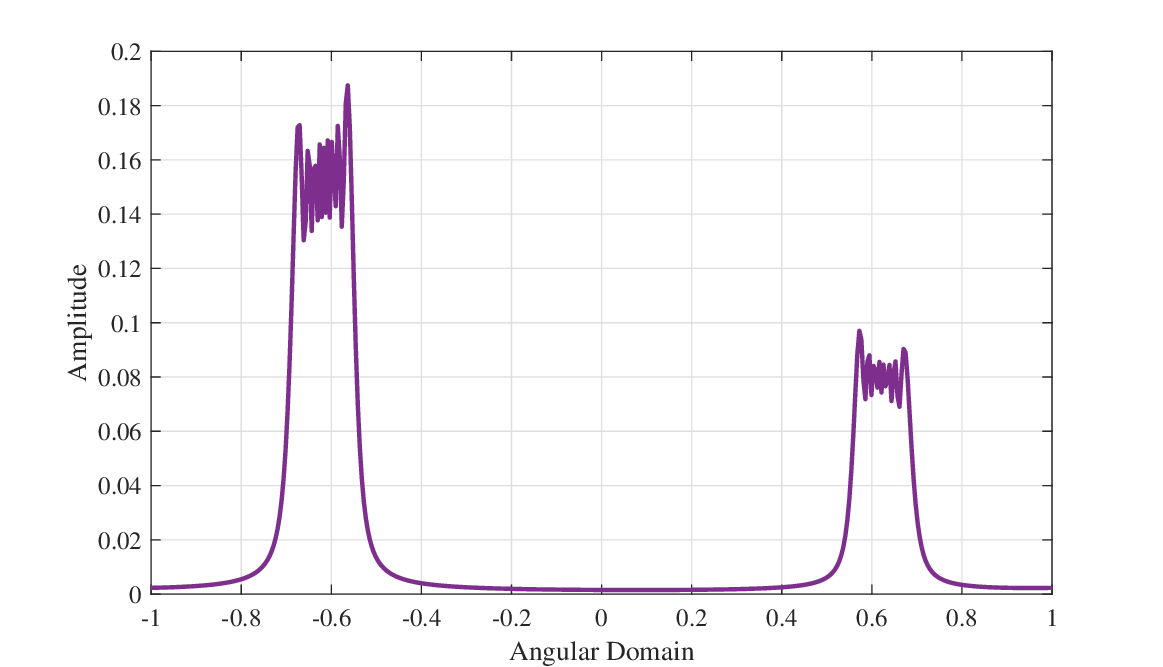}
    \caption{Amplitude of the two-path NF channel in the angular domain. The RIS has $N_\text{R} = 450$ elements and the carrier frequency is 28 [GHz].}
    \label{fig:single_burst}
    \vspace{-0.65cm}
\end{figure}

Note that the phase of each element in the steering vector~\eqref{eq:near_field_channel} is nonlinear to the array index $n$. Thus, the NF channel cannot be described by a single Fourier vector~\cite{cui2022near}. In fact, the amplitudes of NF channel in the angular domain are characterized by blocks of non-zero elements associated with a series of consecutive Fourier vectors, as shown in Fig.~\ref{fig:single_burst}. This work introduces two novel algorithms, BESVR and TVR, to exploit the structured-sparsity of the NF channels in the angular domain.
\section{BESVR Algorithm}
\label{sec:besvr}
\subsection{First Stage of BESVR Algorithm}
Given the sparsity of the NF channel in the angular domain, the main idea of the BESVR algorithm is to identify the $U$ non-zero elements (referred to as peaks) within the angular domain. This can be achieved, for instance, by finding the elements which provide the largest correlation between the received signal $\mathbf{y}_\text{H}$ and the columns of the sensing matrix $\boldsymbol{\Psi}$. 
The exact number of peaks in the channel is unknown. Therefore, $U$ is chosen as an upper bound for the number of peaks. 
After computing the maximum correlation, each candidate peak $p$ is added to the set of possible peaks $\boldsymbol{\omega}$. Following that, we sort the set $\boldsymbol{\omega}$ in an ascending order and compute the initial and final boundaries of the blocks, $\omega_\text{low}$ and $\omega_\text{up}$, respectively. Based on the estimated boundaries, we compute the locations of the blocks, i.e, $\mathcal{R}$.
\subsection{Second Stage of BESVR Algorithm}
At the second stage, we use the information on the locations of the blocks to simplify the NF CE. In such a case, the elements of $\mathbf{z}$ associated with the complement of set $\mathcal{R}$ are set to be zero. Thus, we aim to recover the low-dimensional sub-vector $\mathbf{z}_{\mathcal{R}}\in\C^{|\mathcal{R}| \times 1}$. The problem is formulated as 
\begin{small}
    \begin{equation}
    \label{eq:lasso_PP}
    \hat{\mathbf{z}}_{\mathcal{R}} = \min_{\mathbf{z}_{\mathcal{R}}}~\frac{1}{2}\|\mathbf{y}_\text{H} -\sqrt{P}\boldsymbol{\Psi}_{\mathcal{R}}\mathbf{z}_{\mathcal{R}}\|^{2}_{2} + \eta\|\mathbf{z}_{\mathcal{R}}\|_{1}, 
    \end{equation}
\end{small}
where $\eta$ is the regularization parameter. We adopt the ADMM strategy to solve the problem in~\eqref{eq:lasso_PP}. For simplicity, we omit the term $\sqrt{P}$ in the computation of the ADMM. By introducing the auxiliary variable $\boldsymbol{\zeta}\in\C^{|\mathcal{R}| \times 1}$, the augmented Lagrangian associated with the problem in~\eqref{eq:lasso_PP} can be written as~\cite{boyd2011distributed}
\begin{small}
    \begin{equation}
        \mathcal{L} (\mathbf{z}_{\mathcal{R}},\boldsymbol{\zeta},\boldsymbol{\varphi}) =  \frac{1}{2}\|\mathbf{y}_\text{H} - \boldsymbol{\Psi}_{\mathcal{R}}\mathbf{z}_{\mathcal{R}}\|^{2}_{2} + \eta\|\boldsymbol{\zeta}\|_{1} + \\
     \frac{\epsilon}{2}\|\mathbf{z}_{\mathcal{R}} -\boldsymbol{\zeta} + \frac{\boldsymbol{\varphi}}{\epsilon}\|_{2}^{2} -\frac{\|\boldsymbol{\varphi}\|_{2}^{2}}{2\epsilon},
    \end{equation}
\end{small}
where $\epsilon>0$ is a penalty parameter and $\boldsymbol{\varphi}\in\C^{{|\mathcal{R}|}\times 1}$ is the dual variable. The key idea of ADMM is to solve the optimization problem through sequential updates of $(\mathbf{z}_{\mathcal{R}},\boldsymbol{\zeta},\boldsymbol{\varphi})$ as follows
\begin{small}
    \begin{equation}
    \label{h_up}
    \mathbf{z}_{\mathcal{R}}^{(i+1)} = \min_{\mathbf{z}_{\mathcal{R}}} \frac{1}{2}\|\mathbf{y}_\text{H} - \boldsymbol{\Psi}_{\mathcal{R}}\mathbf{z}_{\mathcal{R}}\|^{2}_{2} + 
    \frac{\epsilon}{2}\|\mathbf{z}_{\mathcal{R}} -\boldsymbol{\zeta}^{(i)} + \frac{\boldsymbol{\varphi}^{(i)}}{\epsilon}\|_{2}^{2},
    \end{equation}
\end{small}
\begin{small}
    \begin{equation}
    \label{eq:zeta_up1}
    \boldsymbol{\zeta}^{(i+1)} = \min_{\boldsymbol{\zeta}}  \epsilon\|\boldsymbol{\zeta}\|_{1} + \frac{\epsilon}{2}\|\mathbf{z}_{\mathcal{R}}^{(i+1)} -\boldsymbol{\zeta} + \frac{\boldsymbol{\varphi}^{(i)}}{\epsilon}\|_{2}^{2},
    \end{equation}
\end{small}
\begin{small}
    \begin{equation}
     \label{u_up}
    \boldsymbol{\varphi}^{(i+1)} = \boldsymbol{\varphi}^{(i)} + \epsilon(\mathbf{z}_{\mathcal{R}}^{(i+1)} - \boldsymbol{\zeta}^{(i+1)}),
    \end{equation}
\end{small}
where the superscript $(i)$ is the ADMM iteration index. The update of $\mathbf{z}_{\mathcal{R}}$ is computed in closed-form expression by setting the gradient of the objective function in in~\eqref{h_up} to zero as
\begin{small}
    \begin{equation}
    \label{eq:h_i+1}
    \mathbf{z}_{\mathcal{R}}^{(i+1)} = (\boldsymbol{\Psi}_{\mathcal{R}}^{\mathsf{H}}\boldsymbol{\Psi}_{\mathcal{R}} + \epsilon\mathbf{I})^{-1}\big[\boldsymbol{\Psi}_{\mathcal{R}}^{\mathsf{H}}\mathbf{y}_\text{H} + \epsilon(\boldsymbol{\zeta}^{(i)}-\boldsymbol{\varphi}^{(i)})\big].
    \end{equation}
\end{small}
Similarly, $\boldsymbol{\zeta}^{(i+1)}$ in~\eqref{eq:zeta_up1} is computed setting the gradient of the objective function with respect to $\boldsymbol{\zeta}$ to zero, resulting in
\begin{small}
    \begin{equation}
    \label{eq:z_i+1}
    \boldsymbol{\zeta}^{(i+1)} = \mathcal{S}(\mathbf{z}_{\mathcal{R}}^{(i+1)} + \boldsymbol{\varphi}^{(i)}), 
    \end{equation}
\end{small}
where $\mathcal{S}$ is the soft thresholding operator defined as $\mathcal{S}(\mathbf{a}) = \text{sign}(\mathbf{a})(|\mathbf{a}|-\frac{\eta}{\epsilon})_{+}$ with the subscript~``$+$'' indicating the positive part of the vector~\cite{donoho1995noising}. The ADMM updates are repeated until the maximum number of iterations is reached, i.e, $i=i_\text{max}$. The estimate of the sub-vector $\mathbf{z}_{\mathcal{R}}$ leads us to reconstruct the transformation coefficient vector $\mathbf{z}$. Thus, we compute $\hat{\mathbf{h}}_\text{MR} = \mathbf{F}\hat{\mathbf{z}}$. 
\textbf{Algorithm 1} summarizes the BESVR approach.
\begin{algorithm}[t]
    \small
    \SetAlgoLined
    \SetKwInput{Input}{Input}
    \SetKwInput{Output}{Output}
    \Input{$\mathbf{y}_\text{H},\boldsymbol{\Psi},U,\eta, \epsilon$}
    \Comment{First Stage} Initialize: $\mathbf{r} = \mathbf{y}_\text{H}$ \\
    \For{$u\to 1$ \KwTo $U$}{
    Compute the maximum correlation between the columns of
    $\boldsymbol{\Psi}$ and $\mathbf{r}$: \\
    $p^{*} = \argmax_{p=1,\cdots,N_\text{R}}|\mathbf{r}^\mathsf{H} [\boldsymbol{\Psi}]_{:,p}|$ \\
    Update the support set: $\boldsymbol{\omega} = \boldsymbol{\omega}\cup p^{*}$ \\
    Compute  $\hat{\mathbf{z}}_{\boldsymbol{\omega}} = \big[(\boldsymbol{\Psi}_{\boldsymbol{\omega}}^{\mathsf{H}}\boldsymbol{\Psi}_{\boldsymbol{\omega}})^{\dagger}\boldsymbol{\Psi}_{\boldsymbol{\omega}}^{\mathsf{H}}\big]\mathbf{y}_\text{H}$ \\
    Calculate the residual: $\mathbf{r} =  \mathbf{y}_\text{H} - \boldsymbol{\Psi}_{\boldsymbol{\omega}}\hat{\mathbf{z}}_{\boldsymbol{\omega}}$ \\
    }
    Sort $\boldsymbol{\omega}\in\mathbb{C}^{U\times 1}$ in ascending order \\
    Get the boundaries: $\omega_\text{low}= \boldsymbol{\omega}_{1}$ and $\omega_\text{up}= \boldsymbol{\omega}_U$\\
    Compute the range of non-zero elements~$\mathcal{R} = \{\omega_\text{low},\cdots,\omega_\text{up}\}$\\
    \Comment{Second Stage} Recover the sub-vector $\mathbf{z}_{\mathcal{R}}$ using~\eqref{u_up},~\eqref{eq:h_i+1}, and~\eqref{eq:z_i+1} \\
    Reconstruct $\hat{\mathbf{z}}$\\
    Get the NF channel vector: $\hat{\mathbf{h}}_\text{MR} = \mathbf{F}\hat{\mathbf{z}}$ \\
    \Output{$\hat{\mathbf{h}}_\text{MR}$}
    \caption{\small BESVR algorithm}
\end{algorithm}
\section{NF CE via Linear TVR}
In this section, we propose an one stage method to recover the NF channel vector. Instead of estimating the locations of the blocks of non-zeros elements, we aim at enforcing the block structure by including a linear TVR penalty in the optimization problem. Thanks to their capability of preserving edges and enforcing smoothness, linear TVR has been used in a variety of applications, such as image processing~\cite{679423}. By incorporating the linear TVR penalty, the optimization can be formulated as
\begin{small}
    \begin{equation}
    \label{eq:flasso}
    \hat{\mathbf{z}} = \min_{\mathbf{z}} \frac{1}{2}\|\mathbf{y}_\text{H} - \sqrt{P}\boldsymbol{\Psi}\mathbf{z}\|^{2}_{2}  + \eta_\text{TV}\sum_{j=2}^{N_\text{R}}|\mathbf{z}_{(j)} - \mathbf{z}_{(j-1)}|, 
    \vspace{-0.2cm}
    \end{equation}
\end{small}
where $\eta_\text{TV}$ is the regularization parameter. Note that the linear TVR computes the absolute difference between two consecutive elements in $\mathbf{z}$. This effectively penalizes variations or abrupt changes in the signal and enforce the block structure.

We employ the ADMM framework to obtain a closed-form solution to the problem presented in~\eqref{eq:flasso}. First, we rewrite the problem in~\eqref{eq:flasso} as~\cite{boyd2011distributed}
\begin{small}
    \begin{equation}
    \label{eq:flasso1}
    \hat{\mathbf{z}} = \min_{\mathbf{z}} \frac{1}{2}\|\mathbf{y}_\text{H} - \sqrt{P}\boldsymbol{\Psi}\mathbf{z}\|^{2}_{2}  + \eta_\text{TV}\|\mathbf{D}\mathbf{z}\|_{1}, 
    \end{equation}
\end{small}
where $\mathbf{D}$ is a matrix of first differences~\cite{hastie}. The problem in the ADMM form is given as
\begin{small}
    \begin{equation}
    \label{eq:flasso2}
    \min_{\mathbf{z},  \boldsymbol{\beta}}\frac{1}{2}\|\mathbf{y}_\text{H} - \sqrt{P}\boldsymbol{\Psi}\mathbf{z}\|^{2}_{2} +  \eta_\text{TV}\|\boldsymbol{\beta}\|_{1}, \mbox{s.t.}~\mathbf{D}\mathbf{z}-\boldsymbol{\beta} = 0
    \end{equation}
\end{small}
where $\boldsymbol{\beta}$ is the auxiliary variable. The computation of the ADMM updates follows similar steps to those detailed in Sec.~\ref{sec:besvr}. Herein, we also omit the term $\sqrt{P}$ in the computations of the ADMM, summarized as follows:
\begin{small}
    \begin{equation}
    \label{eq:z_tv}
    \mathbf{z}^{(k+1)} = (\boldsymbol{\Psi}^{\mathsf{H}}\boldsymbol{\Psi} + \epsilon_\text{TV}\mathbf{D}^{\mathsf{T}}\mathbf{D})^{-1}\big[\boldsymbol{\Psi}^{\mathsf{H}}\mathbf{y}_\text{H} + \epsilon_\text{TV}\mathbf{D}^{\mathsf{T}}(\boldsymbol{\beta}^{(k)}-\boldsymbol{\xi}^{(k)})\big]
    \end{equation}
\end{small}
\begin{small}
    \begin{equation}
    \label{eq:b_tv}
    \boldsymbol{\beta}^{(k+1)} =\mathcal{S}(\mathbf{D}\mathbf{z}^{(k+1)} + \boldsymbol{\xi}^{(k)})
    \end{equation}
\end{small}
\begin{small}
    \begin{equation}
    \label{eq:xi_tv}
    \boldsymbol{\xi}^{(k+1)} = \boldsymbol{\xi}^{(k)} +\epsilon_\text{TV}(\mathbf{D}\mathbf{z}^{(k+1)} - \boldsymbol{\beta}^{(k+1)}),
    \end{equation}
\end{small}
where the superscript $(k)$ denotes the iteration index for this problem, $\boldsymbol{\xi}$ is the dual variable, and $\epsilon_\text{TV}$ is the penalty  factor. The soft-thresholding operator in~\eqref{eq:b_tv} is defined in a similar way to~\eqref{eq:zeta_up1}. Furthermore, the ADMM updates are iterated until reaching the maximum number of iterations, i.e, $k = k_\text{max}$. \textbf{Algorithm 2} outlines the proposed TVR-ADMM algorithm. 
\begin{algorithm}[t]
    \small
    \SetAlgoLined
    \SetKwInput{Input}{Input}
    \SetKwInput{Output}{Output}
    \Input{$\mathbf{y}_\text{H},\boldsymbol{\Psi},\eta_\text{TV},\epsilon_\text{TV}$}
    \textbf{Initialization}: $\mathbf{z}^{(0)}$, $\boldsymbol{\beta}^{(0)}$ and $\boldsymbol{\xi}^{(0)},k=0$ \\
   \Repeat{stopping criterion is satisfied}{
    Update $\mathbf{z}$ using equation~\eqref{eq:z_tv} \\
    Update $\boldsymbol{\beta}$ using equation~\eqref{eq:b_tv} \\
    Update $\boldsymbol{\xi}$ using equation~\eqref{eq:xi_tv} \\
    $k=k+1$
   }
    Compute NF channel vector: $\hat{\mathbf{h}}_\text{MR} = \mathbf{F}\hat{\mathbf{z}}$ \\
    \Output{$\hat{\mathbf{h}}_\text{MR}$}
    \caption{\small Linear TVR-ADMM}
\end{algorithm}
\section{Numerical Results}
We consider $N_{\text{R}} = 450$, $L_{\text{MR}} = \{1, 2\}$ and $f_c = 28$ [GHz]. Unless stated otherwise, the number of RF chains at RIS is set to $N_\text{RF,R} = 5$. The SNR is defined as $\frac{1}{\sigma^{2}}$, and the training overhead is set as $T = 45$ time slots. The analog combining matrix at the hybrid RIS $\mathbf{W}_\text{M}$ is constructed with random phases within $\mathcal{U}[0, 2\pi]$. The upper bound for the number of peaks in the BESVR algorithm is set to $U=30$. The maximum numbers of ADMM iterations are set as $i_\text{max} = |\mathcal{R}|$ and $k_\text{max} = N_\text{R}$. The distances between RIS and MS are randomly selected from $\mathcal{U}[10,50]$ [m] while the angles are sorted within the interval $[-\pi,\pi]$. The regularization parameters $\eta$, $\epsilon$, $\eta_\text{TV}$ and $\epsilon_\text{TV}$ are determined through cross-validation analysis. The performance of the algorithms is measured in terms of NMSE of the channel vector and average CPU time. The simulation results in terms of NMSE are obtained by averaging over 200 independent trials. The performance of our proposed methods is compared with the following baselines: polar domain representation for NF CE (P-OMP)~\cite{cui2022near} and the burst-LASSO~\cite{liu2016exploiting} algorithm.

\begin{figure}[t]
    \centering
    \includegraphics[width=0.88\columnwidth]{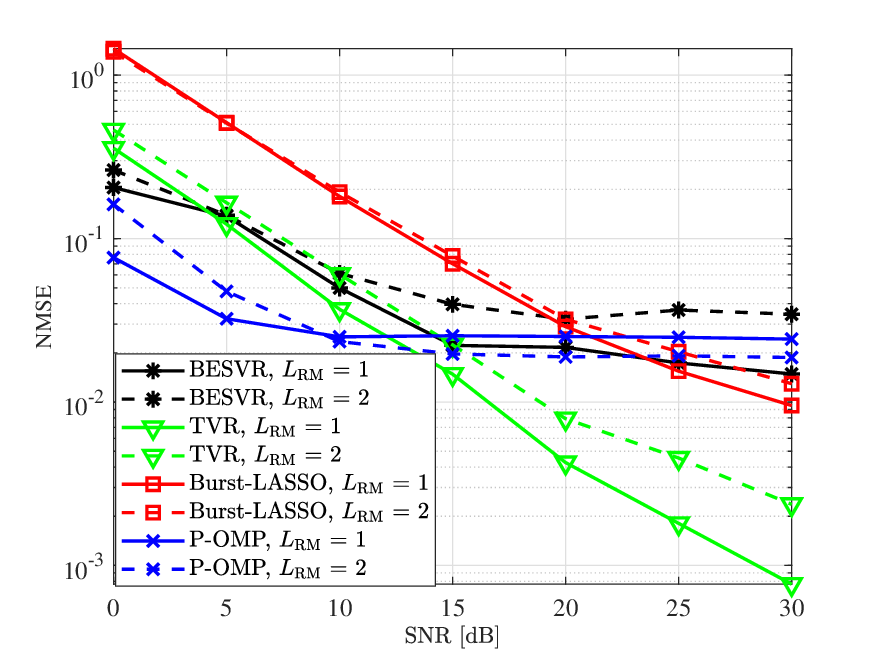}
    \caption{NMSE versus SNR [dB].}
    \label{fig:nmse_snr}
    \vspace{-0.4cm}
\end{figure}
The NMSE performance of the proposed methods against the SNR values is shown in Fig.~\ref{fig:nmse_snr}. Focusing on single-path scenario, we observe that the P-OMP has the lowest NMSE between the estimators at SNR $5$ [dB]. However, in a high SNR regime, the NMSE performance of P-OMP method saturates to the level $10^{-2}$ while the linear TVR algorithm is able to gradually improve its performance. We further notice that the linear TVR algorithm outperforms the burst-LASSO algorithm and the BESVR algorithm. In addition, we see that the BESVR can outperform the P-OMP algorithm when the SNR is set to $20$ [dB]. Now, we extend our analysis to the multi-path scenario. We observe from the figure that the burst-LASSO and the linear TVR algorithms have slightly worse performance in this scenario. However, thanks to their robustness, both algorithms still can outperform the P-OMP in the high SNR regime. Moreover, we notice that the BESVR suffers from performance degradation in the multi-path scenario. In this case, the estimation of the boundaries of the blocks becomes more challenging since there is a higher probability of detecting false alarms instead of the peaks of the blocks.  Hence, the BESVR method could be enhanced for improved CE accuracy.

\begin{figure}
    \centering
    \includegraphics[width=0.88\columnwidth]{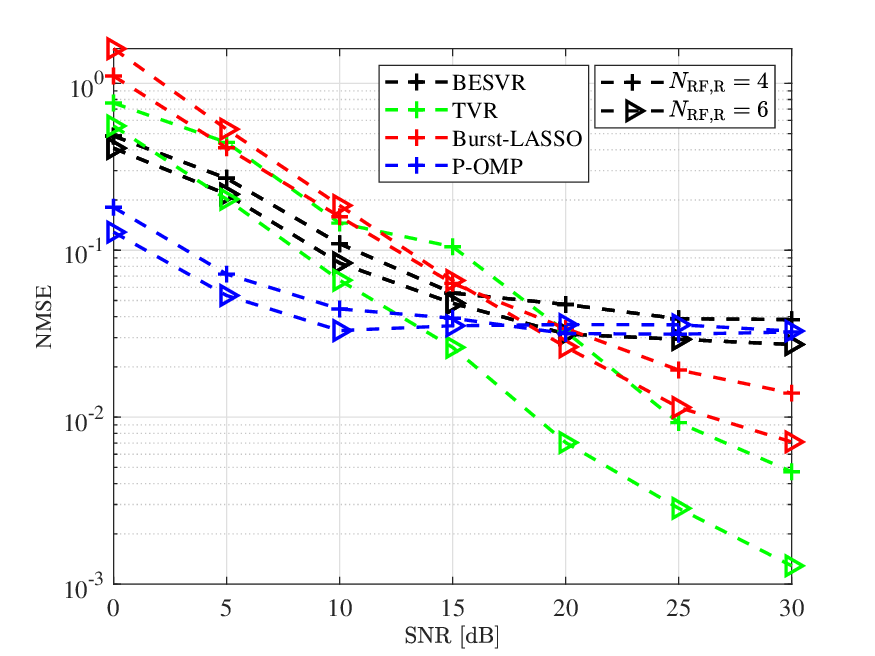}
    \caption{Effect of the number of RF chains at RIS on the NMSE performance.}
    \label{fig:nmse_snr_act}
        \vspace{-0.2cm}
\end{figure}

The effect of the number of RF chains at RIS on the NMSE performance in multi-path scenario is depicted in Fig.~\ref{fig:nmse_snr_act}. For this experiment, we set $N_\text{RF,R} \in \{4,6\}$. Clearly, the number of RF chains at RIS does not affect the P-OMP performance significantly. However, the linear TVR algorithm shows a significant performance gain when the number of RF chains is set to $N_\text{RF,R} = 6$. In other words, the performance of the linear TVR algorithm could be further improved by deploying more RF chains to the RIS. Besides, we notice that the linear TVR algorithm is able to outperform the baseline schemes even with a lower number of RF chains at the RIS, i.e., $N_\text{RF,R} = 4$.
\begin{figure}
    \vspace{-0.3cm}
   \centering
    \includegraphics[width=0.88\columnwidth]{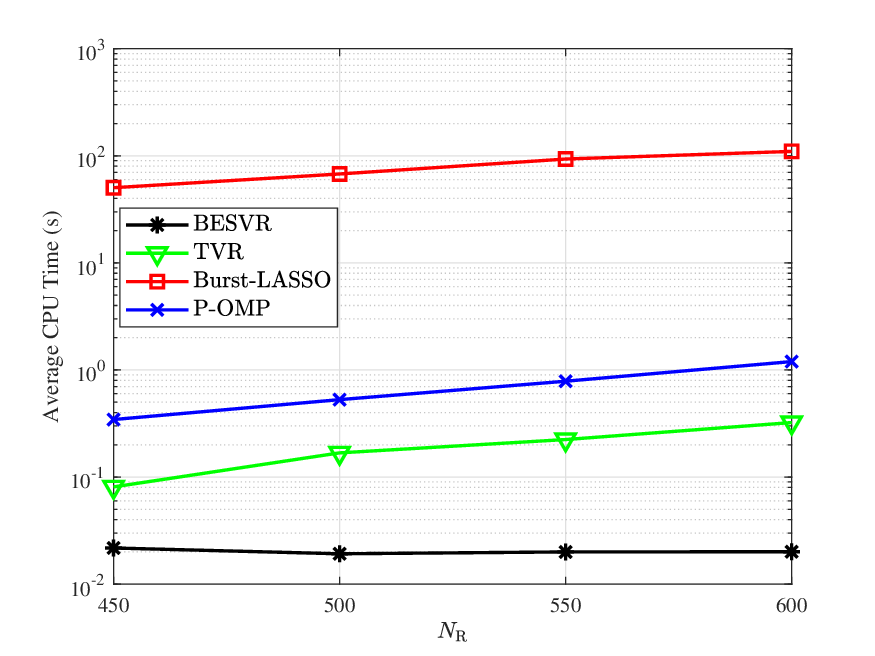}
    \caption{Average CPU time in seconds (s) versus $N_\text{R}$.} 
    \label{fig:time_nr}
    \vspace{-0.6cm}
\end{figure}

Fig.~\ref{fig:time_nr} examines the effect of RIS elements on the average CPU time of the estimators, using $10$ Monte-Carlo simulations for a consistent comparison.
We assume $L_\text{MR} = 1$ and the SNR is set to be $20$ [dB]. The results show that the burst-LASSO algorithm has the highest average CPU time among the estimators.
Besides, we observe that the linear TVR algorithm has a lower CPU time compared with the P-OMP. For instance, when $N_\text{R} = 450$, the average CPU time of the linear TVR algorithm is only $0.08$ (s), while the P-OMP requires $0.34$ (s) to compute one trial. Also, note that the CPU time of the burst-LASSO, P-OMP, and linear TVR algorithms is associated with the number of elements at RIS while the complexity of the BESVR algorithm is associated with the estimated length of the blocks of non-zero elements. Thus, the BESVR algorithm is able to keep the CPU time constant despite the high number of RIS elements.
\section{Conclusions} 
We have proposed two novel algorithms to tackle the NF CE problem: BESVR and linear TVR. Besides, we have developed an ADMM-based algorithm to reduce the computational complexity. Our results have shown the benefits of our proposed methods in terms of CPU time and NMSE performance. 
\section*{Acknowledgements}
The work was supported in part by the Research Council of Finland (former Academy of Finland) 6G Flagship Program (grant 346208) and EERA project (grant 332362).
\balance
\bibliographystyle{IEEEtran}
\bibliography{IEEEabrv,references}
\end{document}